\documentclass[aps,a4paper,12pt,onecolumn,final,notitlepage,oneside]{article}
\usepackage[cp1251]{inputenc}
\usepackage[T2A]{fontenc}
\usepackage[english,russian]{babel}
\usepackage{indentfirst}
\usepackage{geometry}
\usepackage[dvips]{graphicx}
\usepackage[final]{epsfig}
\usepackage{multicol}
\usepackage{graphicx}
\usepackage{subfigure}
\usepackage{amsmath}
\usepackage{enumerate}
\usepackage{amssymb}
\usepackage{amscd}
\usepackage{hhline}
\usepackage{multirow}
\usepackage{setspace}
\usepackage{makeidx}
\usepackage{textcomp}
\usepackage{amsfonts}
\usepackage{afterpage}
\usepackage{longtable}
\usepackage{cite}
\usepackage{rawfonts}
\usepackage{array}
\begin{document}
\title{ On a Special Transformation to a Non-Inertial,
Radially Rigid Reference Frame }
\author{V. V. Voytik\thanks {E-mail: voytik1@yandex.ru,   Bashkir State Pedagogical University, ul. Oktyabr’skoi Revoliutsii 3-a, Ufa 450077, Russia,
Received October 26, 2012}}

\maketitle
\selectlanguage{english}
\begin{abstract}
We discuss the conditions under which a body, moving non-inertially in Minkowski space,
can preserve its size. Under these conditions, using a series expansion of the generalized Lorentz
transformation, we find a coordinate transformation connecting the laboratory inertial reference frame $S$
and the rigid non-inertial reference frame $s$ which moves without its own rotation with respect to $S$. Direct
consequences of this transformation are: (a) desynchronization, in system $s$, of the coordinate clocks of $s$
which were previously synchronized in $S$, and (b) a kinematic contraction of a ruler of system s observed in
$S$. We also consider the dependence of the transformation vector parameter on the proper coordinates of $s$.
\end{abstract}

\subsection*{1. INTRODUCTION}

In special relativity (SR), there exists a special
transformation of the 4-coordinates which has the
meaning of a transition from a laboratory inertial reference
frame $S:(T,\mathbf{R})$ to a radially rigid, arbitrarily
accelerated non-inertial reference frame $s:(t,\mathbf{r})$ (see
Section 2 below). It has initially received the name
of the generalized Lorentz transformation \cite {8}, but
further on it will be called the special Lorentz–Moller-
Nelson (LMN) transformation.

The reference frame obtained with this transformation
possesses the following properties. Its time
coordinate is the proper time of the origin, the spatial
coordinate axes are Cartesian, and the square root
of the sum of squares of the coordinates of a point
is the distance between this point and the origin. \footnote[2]{The 3-space metric of an arbitrary radially rigid nonstationary
reference frame is still variable and non-Euclidean. In
other words, the distance between two fixed points in this
reference frame can change, therefore the rigidity of such a
reference frame is only radial.}
In such 4-coordinates, the metric has an especially
simple form \cite {15}  (see also \cite[p. 404, Eq.(13.71)]{16}), with
the correction that there is no space-time curvature
in SR.

An important role of the LMN transformation in
gravitational field theory is contained in the existence
of a kinematic approach to gravity. By this approach,
an observer at rest in a gravitational field can locally
consider him/herself as if moving with acceleration
in a space without gravitational field. Meanwhile, an observer falling in a gravitational field can be considered
as if being at rest in a laboratory inertial reference
frame. By the equivalence principle (EP), all
observational results will be independent of whether
there is a local gravitational field “as a matter of
fact”. Due to this circumstance, to connect the two
reference frames, the falling one and the one at rest,
one can locally ignore the gravitational field and consider
motions in flat space-time. Another important
meaning of this transformation in physics is that the
explicit coefficients of its differential form represent a
connection between tensors in the laboratory frame
and the moving, radially rigid frame.

The reference frames considered in general relativity
(GR) are usually non-rigid since perfectly rigid
bodies do not exist. \footnote[3]{In relativity, speaking of a perfectly rigid body, one usually
means a fictitious body with conserved distances between a
selected point which moves in a certain way and any other
point, left to itself.} It can therefore seem that
this transformation is useless, or is of methodological
interest only. However, under certain conditions, to
be discussed in Section 3, the radial rigidity of a
reference frame will be a good approximation. In this
case, the LMN transformation is meaningful in GR.

Thus it makes sense to find out the physical consequences
of the LMN transformation and its characteristic
features that distinguish it from the conventional
Lorentz transformation. Let us recall that if $s$
is an inertial reference frame, then, apart from time
slowing-down, there are two other basic kinematic
effects. These are the Fitzgerald-Lorentz length contraction
and Einstein’s relativity of simultaneity. A non-inertial nature of a reference frame, in general,
exerts an influence on these effects, slightly changing
their magnitudes \cite {4}, \cite {12}. Besides, there emerges the
effect of inhomogeneity of motion of points belonging
to an accelerated reference frame \cite {4}. It is also
necessary to take into account that each effect should
be considered in both reference frames $S$ and $s$, since
the results in the relativistic case are different, and,
in particular, for time it is necessary to distinguish
the coordinate desynchronization of clocks from the
physical one.

In \cite {11}, in Eqs. (4) and (13) for an arbitrary reference
frame $s$, two effects were considered: (a) in a
laboratory inertial frame, the desynchronization effect
for coordinate clocks at rest, synchronized in frame
$s$, and (b) the Lorentz length contraction effect for an
$S$ -frame ruler with respect to $s$.

In \cite {4}, \cite {12}, for a rectilinear motion of $s$, four effects
were considered: (a) Lorentz contraction of an $s$-ruler
in $S$; (b) desynchronization of coordinate clocks of $s$
in $s$ if they had been synchronized in $S$; (c) inhomogeneity
of the point motion velocities of the reference
frame $s$, and (d) desynchronization in $S$ of the clocks
measuring the physical time in the accelerated frame.
The inhomogeneity of point motion velocities of $s$ in
the case of a rectilinear motion was also considered
in \cite[p. 65, Eq. (6)] {3}.

The purpose of the present paper is to determine
the first two effects as functions of the proper acceleration,
but now for arbitrary motions of frame $s$,
and to find out the approximate dependence of the
transformation parameter $\mathbf{v}$ on the proper coordinates.
The physical consequences of the special LMN
transformation manifest themselves most clearly if
one finds, beforehand, its inverse transformation. The
form of the special LMN transformation, unlike that
of the Lorentz transformation, is nonlinear, therefore
a search for its inverse transformation is not a simple
task. This transformation will be sought for by consecutive
approximations, being restricted to second
powers of proper coordinates of frame $s$.

Further on, in Sections 3 to 5, we consider step
by step each relativistic effect, and in Section 6 our
resulting relations are compared with those already
known.

\subsection*  {2. THE SPECIAL TRANSFORMATION
TO A RADIALLY RIGID REFERENCE FRAME}

 The transformation from the laboratory inertial
reference frame $S:(T,\mathbf{R})$  to the rigid non-inertial
frame $s:(t,\mathbf{r})$, moving with respect to $S$ without
rotation, has the form \cite {8} (here and henceforth $c=1$)

\begin{equation} \label{10}
     T=\frac{\mathbf{vr}}{\sqrt{1-v^{2} } } +\int _{0}^{t}\frac{dt}{\sqrt{1-v^{2} } }\,,
\end{equation}

\begin{equation} \label{11}
 \mathbf{R=r}+\frac{1-\sqrt{1-v^{2} } }{v^{2} \sqrt{1-v^{2} } } \mathbf{(vr)v}+\int _{0}^{t}\frac{\mathbf{v}dt}{\sqrt{1-v^{2} } }\, .
\end{equation}
Here $T$ and $\mathbf{R}$ are the time and spatial coordinates
of the inertial frame $S$; $t$ and $\mathbf{r}$ are the same for the
non-inertial frame $s$, respectively. This transformation
is parametrized by only one quantity $\mathbf{v}(t)$ (the
time-dependent velocity of frame $s$), therefore such
an orbital motion of $s$ can be called translational. It
turns out, however, that the reference frame $s$ also has
a certain proper rotation, connected in a certain way
with its orbital motion. This rotation is the Thomas
proper precession.

Substituting the differentials of \eqref{10} and \eqref{11} into
the expression for the interval of the inertial frame in
rectilinear coordinates,
 \begin{equation} \label{14}
ds^{2} =dT^{2} -d\mathbf{R}^{2}\, ,
\end{equation}
we obtain the well-known interval \cite{8},  \cite{15}
\begin{equation} \label{3}
ds^{2} =\left[\, (1+\mathbf{Wr})^{2} -(\mathbf{\Omega \, \, \times r})^{2} \right]^{} \, dt^{2}\
 -2(\mathbf{\Omega \times r})d\mathbf{r}\, dt-d\mathbf{r}^{2}\, ,
\end{equation}
for a rigid reference frame with its proper acceleration $\mathbf{W}$  and rotating with the Thomas proper precession
rate $\mathbf{\Omega}$ , where

\begin{equation} \label{15}
 \mathbf{W}=\frac{\mathbf{\dot{v}}}{\sqrt{1-v^{2} } } +\frac{1-\sqrt{1-v^{2} } }{v^{2} (1-v^{2} )} \mathbf{(\dot{v}v)v}\,,
\end{equation}
\begin{equation} \label{16}
\mathbf{\Omega} = \mathbf{\Omega}_{T}=\frac{1-\sqrt{1-v^{2} } }{v^{2} \sqrt{1-v^{2} } }\mathbf{v\times \dot{v}}\,.
\end{equation}
This angular frequency depends on the nature of the
orbital motion of the reference frame. These relations
can be rewritten in terms of the laboratory acceleration
$\mathbf{\dot{V}}={d\mathbf{V}}/{dT}$
\begin{equation} \label{54}
\mathbf{W}=\frac{\mathbf{\dot{V}}}{1-V^{2} } +\frac{1-\sqrt{1-V^{2} } }{V^{2} \sqrt{1-V^{2} } ^{\,\,3} } (\mathbf{\dot{V}V)V}\,,
\end{equation}
\begin{equation} \label{55}
\mathbf{\Omega} _{T} =\frac{1-\sqrt{1-V^{2} } }{V^{2} (1-V^{2} )} \mathbf{V\times \dot{V}}\,.
\end{equation}

From \eqref{54} it follows that
\begin{equation} \label{56}
\mathbf{\dot{V}}=(1-V^{2} )\mathbf{W}
-\frac{(1-V^{2} )(1-\sqrt{1-V^{2} } )}{V^{2} }\mathbf{(VW)V}\,,
\end{equation}
\begin{equation} \label{57}
\mathbf{V\dot{V}}=\sqrt{1-V^{2} } ^{\,\,3} \mathbf{VW}\,.
\end{equation}

\subsection*  {3. THE RIGID BODY APPROXIMATION}

The LMN transformation is only correct for a perfectly
rigid body, whereas a real, arbitrarily moving
reference frame does not preserve its proper rulers.
For example, the proper size of an elastic body, whose certain point is subject to a jump of velocity or acceleration,
will inevitably change. Let us make it clear,
in which case a moving elastic body can be treated as
an almost rigid one in its proper reference frame. This
question is answered by elasticity theory. According
to it, the self-energy of an elastic deformation of a
ruler is a functional of the specific law of motion of the
ruler’s certain point, taken for the origin. A sudden
impact, applied to the origin of a ruler fabricated from
a homogeneous elastic material, launches deformation
waves propagating with a characteristic period
equal to the ratio of the ruler’s proper length $r$ to the
velocity of sound in its material, $v_{s}$. If the proper
acceleration $\mathbf{W}$ is changing slowly enough during
this period, i.e., if

\begin{equation} \label{2}
\frac{{dW}}{{dt}} \ll \frac{{{v_sW}}}{r}
\end{equation}
the proper acceleration of the ruler’s origin can be
considered as an adiabatically slowly changing parameter.
In this case, both the ruler’s free energy and
its proper deformation, averaged over the period, will
depend (apart from the choice of the material) only
parametrically on the proper acceleration value at a
given time instant.

Such an impact on the elastic ruler’s scale, averaged
over the period, since it depends on the material,
can be easily taken into account by simply redenoting
the ruler’s graduations. Apart rom such a
systematic action, the ruler’s coordinate will possess
an error $\delta r$ due to oscillations caused by the inhomogeneous
acceleration. Its value will have the order
of magnitude of a product of the scale graduation’s
velocity $v_{r}$ due to a jerk  
${dW}/{dt}\;$ by the period of the
elastic wave. This velocity will in turn be of the order
\begin{equation} \label{1.1}
{{v}_{r}}\sim \frac{{{r}^{2}}}{{{v}_{s}}^{2}}\frac{dW}{dt}\,.
\end{equation}
Consequently,
\begin{equation} \label{2.6}
\delta r\sim \frac{{{r}^{3}}}{{{v}_{s}}^{3}}\frac{dW}{dt}\ll r\,.
\end{equation}
Hence,
\begin{equation} \label{2.7}
\frac{dW}{dt}\ll \frac{{{v}_{s}}^{3}}{{{r}^{2}}}\,.
\end{equation}
Under the conditions \eqref{2}, \eqref{2.7}, the elastic ruler can
be considered to preserve its size. These requirements
are certainly classical because the ruler of the reference
frame has been assumed to be a classical body,
and the whole reasoning concerned only its proper
reference frame.

The rigid body approximation constrains the required
precision in calculating the quantities characterizing
the non-inertial reference frame. Such
quantities are represented by power series in $r$. If we restore $c$, then the first-order term will be proportional
to ${Wr}/{{{c}^{2}}}\;$. The next terms of the expansion of
the parameter $\mathbf{v}$ in powers of $\mathbf{r}$ will be proportional
to ${{{(Wr)}^{2}}}/{{{c}^{4}}}\;$ and ${{r}^{2}}dW/{{c}^{3}}dt$. For a real rather
than perfect body, there will emerge one more term
connected with an error in determining the coordinate \eqref{2.6} due the ruler’s oscillations. This termwill be
of the order
  ${{{r}^{3}}WdW}/{{{c}^{2}}{{v}_{s}}^{3}dt}\;$.   The strictness of the
conditions \eqref{2} and \eqref{2.7} does not in any way prevent
one to choose the quantity $dW/{dt}\;$ to be of the order
  \begin{equation} \label{15.1}
\frac{dW}{dt}\sim \frac{{{W}^{2}}}{c}
\end{equation}
 or
  \begin{equation} \label{15.2}
  \frac{dW}{dt}\sim \frac{{{v}_{s}}^{3}}{{{c}^{2}}r}W\,.
  \end{equation}
  In this case the terms with ${{r}^{2}}dW/{{c}^{3}}dt$ and with ${{{r}^{3}}WdW}/{{{c}^{2}}{{v}_{s}}^{3}dt}\;$, respectively, will be of the same
order as ${{{(Wr)}^{2}}}/{{{c}^{4}}}\;$. Therefore, to neglect the corrections
caused by the elastic oscillations (i.e., by the
non-rigidity of the reference frame), it is necessary to
restrict oneself in the expansions to terms containing
the acceleration (but not its derivatives) only linearly.

  A separate problem is to calculate the velocities of
points of an almost rigid reference frame. Then, for
an elastic body, the velocity expansion contains an
additional periodic term of the order ${{r}^{2}}dW/c{{v}_{s}}^{2}dt$ \eqref{1.1}, for which nothing prevents one from choosing
  \begin{equation} \label{15.3}
  \frac{dW}{dt}\sim \frac{{{v}_{s}}^{2}}{{{c}}r}W\,.
  \end{equation}
  In this case such a correction will be of the same
order as ${Wr}/{{{c}^{2}}}\;$. However, if one averages the velocity
expression over the elastic wave period, this averaged
correction vanishes. Thus, in the expansion for the
velocities of points of the reference frame, all quantities
should be understood as being time-averaged. In
what follows, the quantity ${{Wr}\mathord{\left/
 {\vphantom {{Wr} {{c^2}}}} \right.
 \kern-\nulldelimiterspace} {{c^2}}}$ will be considered
to be very small but non-zero. Its higher powers will
be neglected
 \begin{equation} \label{15.6}
{{(Wr)}^{2}}/{{c}^{4}}\cong 0\,.
  \end{equation}

Thus, let a point of a rigid body, taken for the origin,
move sufficiently smoothly, without abrupt jerks
in its proper acceleration. Then the elastic properties
of this body forming the reference frame can be ignored,
and one can consider its proper size to preserve
in the process of motion. In this case, the transformation \eqref{10}, \eqref{11} to a radially rigid non-inertial reference
frame moving with its proper Thomas precession, is
physically meaningful.

 \subsection*  {4. THE TRANSFORMATION PARAMETER
AS A FUNCTION OF LABORATORY TIME
AND PROPER COORDINATES}

    The inverse transformation of \eqref{10}, \eqref{11} can be obtained
in the general case in an analytic form only
approximately, as a power series in the components
of $\mathbf{r}$. In what follows, the dependence of a function
on a certain quantity is shown by a subscript. Let us
denote the time integral as

\begin{equation} \label{27}
\int _{0}^{t}\frac{dt}{\sqrt{1-v^{2} } }  =\theta _{t}\,.
\end{equation}
Then, expressing $t$ from \eqref{27}  and substituting it into $\mathbf{v}(t)$,   one can present it as a function of  $\theta $
\begin{equation} \label{29}
\mathbf{v=V}_{\theta }\,.
\end{equation}
So the first equation of the transformation \eqref{10},\eqref{11} takes the form

\begin{equation} \label{30}
T=\theta +\frac{\mathbf{V}_{\theta }\mathbf{r}}{\sqrt{1-V_{\theta } ^{2} } }\,.
\end{equation}

 The function $\mathbf{V}_{\theta } $   is the velocity of a point $\bf{r}$ of the
frame $s$ at time $\theta $   with respect to the inertial reference
frame $S$.
This instant $\theta $ is different for different points
of frame $s$. Let us make clear how it is related to $T$ and $\mathbf{r}$. To solve this equation, we shall use a power
expansion in $\mathbf{r}$ . In the first approximation,  $\theta $  is
\begin{equation} \label{34}
\theta =T-\frac{\mathbf{Vr}}{\sqrt{1-V^{2} } } \,
\end{equation}
where $\mathbf{V=V}_{T} $ is the velocity of the origin of frame $s$ as a function of the laboratory time $T$.

Therefore, in the first approximation in powers of $\mathbf{r}$, the parameter $\mathbf{v}$ is
\begin{equation} \label{37}
\mathbf{v=V}_{\theta}=\mathbf{V}-\frac{\mathbf{(Vr)\dot{V}}}{\sqrt{1-V^2}}\ , \,\,\,\,   \mathbf{\dot{V}}=\frac{d\mathbf{V}}{dT}\, .
\end{equation}
Substituting here Eq. \eqref{56}, we obtain another representation
of this equation:
\begin{equation} \label{38.3}
\mathbf{v}=\mathbf{V}-\sqrt{1-{{V}^{2}}}(\mathbf{Vr})\mathbf{W}
+\frac{\sqrt{1-{{V}^{2}}}(1-\sqrt{1-{{V}^{2}}})}{{{V}^{2}}}(\mathbf{Vr})(\mathbf{VW})\mathbf{V}\,.
\end{equation}
Thus the parameter  $\mathbf{v}$ depends, in addition to laboratory
time $T$, on the point position $\mathbf{r}$ in the coordinate
frame, i.e., it is inhomogeneous. From Eq. \eqref{37} one
can see that to obtain an invariable position of some
point of the coordinate system in a reference frame
(that is, to obtain a rigid reference frame), it is necessary
that this point move in the laboratory inertial
frame in a certain agreement, or correlation, with the
motion of the origin.

 Let us now find the function ${\theta}(T,\mathbf{r})$ in the second
approximation. Expanding ${\mathbf{V}_{\theta }  \mathord{\left/{\vphantom{V_{\theta }  \sqrt{1-V_{\theta }^{2} } }}\right.\kern-\nulldelimiterspace} \sqrt{1-V_{\theta }^{2} } } $ in a Taylor
series and leaving only the first term, we obtain
\begin{equation} \label{35}
\frac{\mathbf{v}}{\sqrt{1-v^{2} } } =\frac{\mathbf{V}_{\theta}}{\sqrt{1-V^{2}_{\theta} } }=\frac{\mathbf{V}}{\sqrt{1-V^{2} } } 
-\frac{\mathbf{(Vr)\dot{V}}}{1-V^{2} } -\frac{(\mathbf{\dot{V}V)(Vr)V}}{(1-V^{2} )^{2} }\, .
\end{equation}

Substituting the expansion \eqref{35} into \eqref{30}, we obtain,
omitting the subscript showing the $T$ dependence
of quantities, in the second-order approximation:
\begin{equation} \label{38}
\theta =T-\frac{\mathbf{Vr}}{\sqrt{1-V^{2} } } +\frac{\mathbf{(Vr)(\dot{V}r)}}{1-V^{2} } +\frac{\mathbf{(\dot{V}V)(Vr)}^{2} }{(1-V^{2} )^{2} } \,.
\end{equation}

\subsection*  {5. NONLINEAR CONTRACTION
OF THE COORDINATE RULER}

Let us denote the time integral in Eq. \eqref{11} as
\begin{equation} \label{28}
\int _{0}^{t}\frac{\mathbf{v}dt}{\sqrt{1-v^{2} } }=\boldsymbol{\lambda} _{t} .
\end{equation}
Then, substituting into
$\boldsymbol{\lambda} _{t} $ the dependence $t_{\theta} $ found
from \eqref{27}, one can, similarly to \eqref{29}, present it as a
function of $\theta $
\begin{equation} \label{29.2}
  \boldsymbol{\lambda}_{t} =\mathbf{\Lambda}_{\theta }\, .
\end{equation}
Then Eq. \eqref{11} takes the form
\begin{equation} \label{31}
\mathbf{R}={\mathbf{\Lambda}_\theta } +\mathbf{r}+\frac{1-\sqrt{1-V_{\theta }^{2} } }{V_{\theta }^{2} \sqrt{1-V_{\theta }^{2} } } (\mathbf{rV}_{\theta } )\mathbf{V}_{\theta } \,,
\end{equation}
where ${\mathbf{\Lambda}_\theta }$ is
\begin{equation} \label{32}
\mathbf{\Lambda}_{\theta } =\int _{0}^{\theta }\mathbf{V}_{\theta } d\theta \, .
\end{equation}

From Eq. \eqref{38} it is evident that an arbitrary vector $\mathbf{x}_{\theta } $ is equal to
\begin{equation} \label{40}
\mathbf{x}_{\theta}=\mathbf{x}-\dot{\mathbf{x}}\frac{\mathbf{Vr}}{\sqrt{1-V^2}}
+\dot{\mathbf{x}}\frac{\mathbf{(Vr)(\dot{V}r)}}{1-V^2}
+\dot{\mathbf{x}}\frac{(\mathbf{\dot{V}V)(Vr)}^{2} }{(1-V^{2} )^{2} }+\frac{1}{2} \ddot{\mathbf{x}}\frac{\mathbf{(Vr)}^{2} }{1-V^{2}}\,,
\end{equation}
where
$$\mathbf{\dot{\mathbf{x}}}=\frac{d\mathbf{x}}{dT} ,  \ddot{\mathbf{x}}=\frac{d^{2} \mathbf{x}}{dT^{2} }$$
, and all quantities
in the r.h.s. of the equalities are taken at time $T$. As an arbitrary vector $\mathbf{x}_{\theta } $ in \eqref{40}    one can take, for
instance, the vector  $\mathbf{\Lambda}_{\theta}$ from \eqref{32}. Hence one can write, omitting the subscript showing the T dependence $T$ ($\mathbf{\Lambda}=\int _{0}^{T}\mathbf{V}dT $)
\begin{equation} \label{42}
\boldsymbol{\lambda}_t =\mathbf{\Lambda}_{\theta } =\mathbf{\Lambda}-\frac{\mathbf{(Vr)V}}{\sqrt{1-V^{2} } } +\frac{\mathbf{(Vr)(\dot{V}r)V}}{1-V^{2} }
+\frac{\mathbf{(Vr)}^{2} (\mathbf{\dot{V}V)V}}{(1-V^{2})^{2}}
+\frac{\mathbf{(Vr)}^{2} \dot{\mathbf{V}}}{2(1-V^{2} )} \, .
\end{equation}

Substituting \eqref{42} into \eqref{31} and using that
\[\frac{1-\sqrt{1-v^{2} } }{v^{2} \sqrt{1-v^{2} } } v_{\alpha } v_{\beta } =\frac{1-\sqrt{1-V^{2} } }{V^{2} \sqrt{1-V^{2} } } V_{\alpha } V_{\beta }
 -\frac{1-\sqrt{1-V^{2} } }{V^{2} (1-V^{2} )} \mathbf{(Vr)}(\dot{V}_{\alpha } V_{\beta } +V_{\alpha } \dot{V}_{\beta } )\]
\begin{equation} \label{44}
-\frac{(1-\sqrt{1-V^{2} } )^2(1+2\sqrt{1-V^{2} } )}{V^{4} (1-V^{2} )^{2} }
 \cdot\mathbf{(\dot{V}V)(Vr)}V_{\alpha } V_{\beta }
\end{equation}
after some transformations we obtain that
\[\mathbf{L=r}-\frac{1-\sqrt{1-V^{2} } }{V^{2} }\mathbf{(Vr)V}
+\frac{1-\sqrt{1-V^{2} } }{V^{2} \sqrt{1-V^{2} } }\mathbf{(Vr)(\dot{V}r)V}\]
\begin{equation} \label{45}
-\frac{(1-\sqrt{1-V^{2} } )^{2} }{2V^{2} (1-V^{2} )} \mathbf{(Vr)}^{2} \mathbf{\dot{V}}
+\frac{(1-\sqrt{1-V^{2} } )^{2} }{V^{4} (1-V^{2} )} \mathbf{(Vr)}^{2} \mathbf{(\dot{V}V)V},
\end{equation}
where
\[\mathbf{L}=\mathbf{R}-\int _{0}^{T}\mathbf{V}dT\,.\]

Here $\mathbf{L}$ is the length of a perfectly rigid rod at time $T$ in the laboratory inertial reference frame $S$, whose
proper length is $\mathbf{r}$ and whose origin moves relative to $S$ with the velocity $\mathbf{V}$ and acceleration $\dot{\mathbf{V}}$. According
to this equation, a sufficiently long straight rod of
length $\mathbf{r}$ looks curved in $S$. Let us solve the resulting
vector equation \eqref{45} with respect to $\mathbf{r}$. To this end, we
multiply both parts of \eqref{45} scalarly by $\mathbf{V}$ and by $\mathbf{\dot{V}}$. We obtain:
 \begin{equation} \label{46}
 \mathbf{LV}=\sqrt{1-V^{2} }\mathbf{Vr}+\frac{1-\sqrt{1-V^{2} } }{\sqrt{1-V^{2} } } \mathbf{(Vr)(\dot{V}r)}
+\frac{(1-\sqrt{1-V^{2} } )^{2} }{2V^{2} (1-V^{2} )} \mathbf{(Vr)}^{2} \mathbf{(\dot{V}V)}\,,
\end{equation}
\[\mathbf{L\dot{V}=\dot{V}r}-\frac{1-\sqrt{1-V^{2} } }{V^{2} } \mathbf{(Vr)(\dot{V}V)}
+\frac{1-\sqrt{1-V^{2} } }{V^{2} \sqrt{1-V^{2} } } \mathbf{(Vr)(\dot{V}r)(\dot{V}V)}\]
\begin{equation} \label{47}
-\frac{(1-\sqrt{1-V^{2} } )^{2} }{2V^{2} (1-V^{2} )} \mathbf{(Vr)}^{2} \mathbf{\dot{V}}^{2}
+\frac{(1-\sqrt{1-V^{2} } )^{2} }{V^{4} (1-V^{2} )} \mathbf{(Vr)}^{2} \mathbf{(\dot{V}V)}^{2}\, .
\end{equation}
Furthermore, we note that it is sufficient to calculate
the factors $\mathbf{\dot{V}r}$ and $\mathbf{Vr}$ in the third, fourth and fifth
terms of the r.h.s. of Eq. \eqref{45} by leaving only the first
power of $\mathbf{L}$. Therefore \eqref{46} implies
\begin{equation} \label{48}
\mathbf{Vr}=\frac{\mathbf{LV}}{\sqrt{1-V^{2} } } \,.
\end{equation}
With the same accuracy, from \eqref{47} with \eqref{48} it follows
\begin{equation} \label{49}
\mathbf{\dot{V}r=L\dot{V}}+\frac{1-\sqrt{1-V^{2} } }{V^{2} \sqrt{1-V^2}} \mathbf{(LV)\dot{V}V}\,.
\end{equation}
The factor $\mathbf{Vr}$ in the second term of \eqref{45} should be
calculated up to the second power of $\mathbf{L}$. Due to the
equalities \eqref{48} \eqref{49}, we obtain from \eqref{46}, that
\[\mathbf{Vr}=\frac{\mathbf{LV}}{\sqrt{1-V^{2} } } -\frac{1-\sqrt{1-V^{2} } }{\sqrt{1-V^{2} } ^{\,\,3} } \mathbf{(LV)(L\dot{V}})\]
\begin{equation} \label{50}
-\frac{(1-\sqrt{1-V^{2} } )^{2} (1+2\sqrt{1-V^{2} } )}{2V^{2} \sqrt{1-V^{2} } ^{\,\,5} } \mathbf{(LV)}^{2} (\mathbf{\dot{V}V})\,.
\end{equation}
Now we substitute \eqref{50} into \eqref{45}, then, using \eqref{48}, \eqref{49} we obtain up to $O(\mathbf{L}^2)$ inclusive that
\[\mathbf{r=L}+\frac{1-\sqrt{1-V^{2} } }{V^{2} \sqrt{1-V^{2} } } \mathbf{(LV)V}
-\frac{(1-\sqrt{1-V^{2} } )^{2} (1+3\sqrt{1-V^{2} } )}{2V^{4} \sqrt{1-V^{2} } ^{\,\,5} } (\mathbf{\dot{V}V)(LV)}^{2} \mathbf{V}\]
\begin{equation} \label{51}
+\frac{(1-\sqrt{1-V^{2} } )^{2} }{2V^{2} (1-V^{2} )^{2} } \mathbf{(LV)}^{2}\mathbf{\dot{V}}
-\frac{1-\sqrt{1-V^{2} } }{V^{2} \sqrt{1-V^{2} } ^{\,\,3} } \mathbf{(LV)(L\dot{V})V}=invariant.
\end{equation}
Eqs. \eqref{45} and \eqref{51} can be expressed in terms of the
proper acceleration. Substituting \eqref{56}, we obtain
\[\mathbf{L=r}-\frac{1-\sqrt{1-V^{2} } }{V^{2} } \mathbf{(Vr)V}
+\frac{(1-\sqrt{1-V^{2}})\sqrt{1-V^2}} {V^{2} }  \mathbf{(Vr)(Wr)V}\]
\begin{equation} \label{59}
-\frac{(1-\sqrt{1-V^{2} } )^{2} }{2V^{2} }\mathbf{(Vr)}^{2}\mathbf{W}
+\frac{(1-\sqrt{1-V^{2} } )^{3} }{2V^{4} }\mathbf{(Vr)}^{2}\mathbf{(VW)V}\,,
\end{equation}
\[\mathbf{r=L}+\frac{1-\sqrt{1-V^{2} } }{V^{2} \sqrt{1-V^{2} } }\mathbf{(LV)V}
-\frac{1-\sqrt{1-V^{2} } }{V^{2} \sqrt{1-V^{2} } } \mathbf{(LV)(LW)V}+\]
\begin{equation} \label{61}
+\frac{(1-\sqrt{1-V^{2} } )^{2} }{2V^{2} (1-V^{2} )} \mathbf{(LV)}^{2} \mathbf{W}
- \frac{(1-\sqrt{1-V^{2} } )^{2} }{V^{4} (1-V^{2} )} \mathbf{(VW)(LV)}^{2}\mathbf{V}\,.
\end{equation}

The nonlinearity of the Lorentz contraction means
that, in general, it is impossible to simulate an arbitrary
reference frame $s$, sufficiently large in size
from the viewpoint of a laboratory observer, by a
sequence of instantaneously comoving inertial reference
frames. Such simulation is only possible for an
observer of frame $s$. This nonlinearity is well explained
by the fact that the points of a rigid reference frame
move in different ways. The region of the coordinate
axis of frame $s$, adjacent to its front end moves, at a
rectilinear acceleration of $s$ with a smaller velocity
than the velocity $V$ of the origin. Therefore, the
region of the coordinate axis close to its front end with
respect to the lab reference frame $S$ is less contracted
that the neighborhood of the origin. Thus the whole
length of the accelerated rod in the process of acceleration
along its own direction should be slightly larger
than the length of a similar but inertial rod, whose all
points move with the same velocity $V$ at a given time
instant \cite{4}.

Eqs. \eqref{45} or \eqref{51} are different forms of one of the
equations of the sought-for inverse transformation,
up to the second order in $\mathbf{r}$  (or $\mathbf{L}$) inclusive. It is
evident that the direct and inverse LMN transformations
are different from each other. This feature is
connected with a radical difference between inertial
and non-inertial reference frames.

\subsection*{6. DESYNCHRONIZATION OF COORDINATE
CLOCKS}

Similarly to \eqref{40}, expanding an arbitrary scalar
function   $y_{\theta } $  in Taylor series up to  $O(\mathbf{r}^2)$      inclusive, we
obtain
\begin{equation} \label{39}
y_{\theta } =y-\left(\frac{\mathbf{Vr}}{\sqrt{1-V^{2} } } -\frac{(\mathbf{Vr)(\dot{V}r)}}{1-V^{2} }\right)\dot{y}
+\frac{(\mathbf{Vr})^{2} (\mathbf{\dot{V}V})}{(1-V^{2} )^{2}} \dot{y}+\frac{(\mathbf{Vr})^{2} }{2(1-V^{2} )} \ddot{y}\,,
\end{equation}
where
$\dot{y}=dy/dT$ and $\ddot{y}=d^{2} y/dT^{2}\, .$
The equation
inverse to \eqref{27}  is
\begin{equation} \label{33}
t=\int _{0}^{\theta }\sqrt{1-V_{\theta } ^{2} } d\theta\,  ,
\end{equation}
where $\theta $   is a root of Eq. \eqref{30}. Substituting into \eqref{39} instead of the scalar function $y_{\theta } $ the function $t_{\theta } $ from \eqref{33}, we obtain
\begin{equation} \label{43}
t=\int _{0}^{T}\sqrt{1-V^{2} } dT
-\left\{\mathbf{Vr}-\frac{\mathbf{(Vr)(\dot{V}r)}}{\sqrt{1-V^{2} } } -\frac{\mathbf{(Vr)}^{2} (\mathbf{\dot{V}V)}}{2\sqrt{1-V^{2} } ^{3} } \right\} \,.
\end{equation}
It is the second equation of the inverse transformation.
The first term in the r.h.s. of this expression
is the proper time of the origin, $\tau$ ($\mathbf{r}=0$). Evidently,
the world time $t$ Evidently,
the world time t cannot coincide with the proper time
of the origin due to relativity of simultaneity. The
second term in curly brackets in \eqref{43} determines the
desynchronization of two coordinate clocks in the
non-inertial frame $s$, if these clocks were initially synchronized
in the lab frame $S$ ($dT=0$). From \eqref{43} it is clear that the desynchronization magnitude, in
the present approximation, does not depend on the
law of motion of frame $s$, but is determined by the
instantaneous values of its velocity and acceleration.

Substituting into \eqref{43} the values $\mathbf{\dot{V}r}$ and $\mathbf{Vr}$ from \eqref{49} and \eqref{50}, we finally obtain another form of
the second equation,
\begin{equation} \label{52}
t=\int _{0}^{T}\sqrt{1-V^{2} } dT
-\left\{\frac{\mathbf{LV}}{\sqrt{1-V^{2} } } -\frac{\mathbf{(LV)(L\dot{V})}}{\sqrt{1-V^{2} } ^{\,\,3} } -\frac{\mathbf{(LV)}^{2} \mathbf{(\dot{V}V)}}{\sqrt{1-V^{2} } ^{\,\,5} } \right\}\,.
\end{equation}
Substituting the equality \eqref{56} into \eqref{43} and \eqref{52}, we
obtain
\begin{equation} \label{58}
t=\int_{0}^{T}\sqrt{1-V^{2}}dT-\mathbf{Vr}+\sqrt{1-V^{2} }\mathbf{(Vr)(Wr)}
+\frac{(1-\sqrt{1-V^2})^2}{2V^2}(\mathbf{Vr})^{2} (\mathbf{VW})\,,
\end{equation}
\begin{equation} \label{60}
t=\int _{0}^{T}\sqrt{1-V^{2} } dT -\frac{\mathbf{LV}}{\sqrt{1-V^{2} } } +\frac{\mathbf{(LV)(LW)}}{\sqrt{1-V^{2} } }
+\frac{(1-\sqrt{1-V^{2} } )}{V^{2} (1-V^{2} )} \mathbf{(LV)}^{2} \mathbf{(VW)}\,.
\end{equation}

Lastly, substituting into \eqref{37} the relation \eqref{48}, we
obtain with the same accuracy
\begin{equation} \label{53}
\mathbf{v=V}-\frac{\mathbf{(VL)\dot{V}}}{1-V^{2} } \,.
\end{equation}

 \subsection*{7. COMPARISON OF THE OBTAINED
FORMULAS WITH KNOWN RESULTS}

First of all, it is necessary to make clear whether
Eqs. \eqref{54} and \eqref{55} are true. Eq. \eqref{54} is known for long time \cite[p. 109, Eq. (194)]{33}. Eq. \eqref{55} also well
agrees with quite a number of other sources, e.g., \cite[Eq.(33)]{38} and \cite[Eq. (34)]{34}.

It is easy to notice that in the case of a constant
velocity $\mathbf{V}$ the inverse LMN transformation \eqref{45}, \eqref{43} and \eqref{51}, \eqref{52} turn into the conventional Lorentz
transformation.

Let us now consider the easily verifiable consequences
of the above inverse transformation in the
case of a rectilinearmotion of frame $s$ without rotation
in the direction of the $X$ and $x$ axes. In this case,
the direction of the laboratory acceleration coincides
with that of the velocity. Then one can verify that
Eqs. \eqref{37}, \eqref{38.3} and \eqref{53} are in agreement with the
already known formulas for the velocities of points of
the coordinate system of $s$ \cite[p. 65, Eq. (6)]{3}, \cite[Eq. (19)]{4}.

Suppose now that two events have occurred at the
same time according to the clocks of the lab inertial
reference frame $S$, but at different points $x_{1} $ and $x_{2} $   of the accelerated reference frame $s$. Let us find the
world time interval between events 1 and 2 in the
accelerated frame $s$. From Eq. \eqref{58} it follows that the
time instants $t_{1} $ and $t_{2}$ by the clocks of frame $s$ are
\begin{equation} \label{68}
t_{\alpha } =\int _{0}^{T}\sqrt{1-V^{2} } dT
-\; \left\{Vx_{\alpha } -V(1-\frac{V^{2} }{2} )\; Wx_{\alpha }^{2}\right\} \,,       (\alpha =1,\; 2 )\,.
\end{equation}
Consequently,
\begin{equation} \label{69}
\Delta t=t_{2} -t_{1} =-\left(Vx-V(1-\frac{V^{2} }{2} )\, Wx^{2} \right)\mathop{{ \mathord{\left/{\vphantom{ }}\right.\kern-\nulldelimiterspace} } }\limits_{x=x_{1} }^{x=x_{2} }\,.
\end{equation}
Similarly, from Eq. \eqref{59} it follows that the length of a
rod of proper length $x$,  situated along the direction of
its rectilinear motion, is
\begin{equation} \label{70}
L=\sqrt{1-V^{2} } x+\frac{V^{2} \sqrt{1-V^{2} } }{2} Wx^{2}\,.
\end{equation}

Eqs. \eqref{69} and \eqref{70}  are already known from a
direct calculation using the equation of rectilinear
uniformly accelerated motion (\cite[Eq. (18)]{12} and \cite[Eq.(26)]{40}, \cite[Eq. (30)]{4}, respectively).
Eqs. \eqref{45} and \eqref{51}, though contradicting to the conventional
Lorentz contraction, still well agree with \cite{11}.

\subsection*{8. CONCLUSION}

 Summarizing, one can make the general conclusion
that the special LMN transformation and the
inverse transformation suggested here are useful for
a description of real, sufficiently rigid bodies. The
physical consequences of the LMN transformation
are nontrivial, they agree with each other and with the
results obtained by other authors.

\subsection*{9. DISCUSSION}

The LMN transformation with the arbitrary parameter $\mathbf{v}(t)$ describes an arbitrary non-inertial rigid
motion of a perfectly rigid body which does not exist.
However, under the constraints \eqref{2}, \eqref{2.7} and for a
reference frame not too large in size, \eqref{15.6}, the idea of
rigid motion in relativity theory is quite meaningful.

 The parameter  $\mathbf{v}\, (t)$ in the LMN transformation
is connected with the velocity of the origin by the
general approximate formula \eqref{37}. This means that
the points of a radially rigid non-inertial frame move
with respect to $S$ (on the average over the oscillation
period) inhomogeneously, with different velocities.

The equations that represent the inverse special
LMN transformation (in two forms, \eqref{45}, \eqref{43} and \eqref{51}, \eqref{52}), respectively) possess an essential nonlinearity
depending on the acceleration of the origin.
The direct and explicit kinematic consequences of
this transformation are: (a) desynchronization (in
Eqs. \eqref{43}, \eqref{52} it is shown in curly brackets) of coordinate
clocks of the non-inertial frame $s$  if they were
previously synchronized in the lab frame $S$, and (b) a
nonlinear contraction in $S$ of a ruler of frame $s$ \eqref{45}, \eqref{51}. In the second order with respect to the proper
coordinates, these effects depend on the velocity and
proper acceleration only. This transformation may be
used for calculating other effects which are possible
at rigid non-inertial motion.

In conclusion, let us briefly dwell on the perspectives
of future studies. The present paper was devoted
to a remarkable transformation from a laboratory
inertial frame to a radially rigid reference frame
which moves arbitrarily but rotates with a strictly
determined Thomas frequency. This proper rotation
is a special case of an arbitrary rotation, therefore
the above formulas of the inverse special LMN
transformation can be generalized. Another direction
of research is contained in the following. The
above transformation is applicable to curved spacetime
only locally, in quite a restricted region of space.
This circumstance is a shortcoming. It is therefore
required to extend the special LMN transformation
to curved space-time. One should expect that the
transformation to be found will be similar to the LMN
transformation but will contain the curvature tensor.
It is quite probable that this more general transformation
will have a wider applicability area.

\subsection*{ACKNOWLEDGMENT}

The author is greatly thankful to Prof. N.G. Migranov
for useful discussions and support.

\newpage
%
%
\end {document}